\documentclass[aps,prl,reprint,superscriptaddress,floatfix,amsmath]{revtex4-1}
\usepackage{graphicx,epstopdf}
\begin{document}

\title{High-precision mass measurement of $^{56}$Cu and the redirection of the rp-process flow}

\author{A.A. Valverde}
\email[]{avalverd@nd.edu}
\affiliation{Department of Physics, University of Notre Dame, Notre Dame, Indiana 46556, USA}

\author{M. Brodeur}
\affiliation{Department of Physics, University of Notre Dame, Notre Dame, Indiana 46556, USA}

\author{G. Bollen}
\affiliation{Department of Physics and Astronomy, Michigan State University, East Lansing, Michigan 48824, USA}
\affiliation{Facility for Rare Isotope Beams, East Lansing, Michigan 48824, USA}

\author{M. Eibach}
\affiliation{Institut f\"{u}r Physik, Universit\"{a}t Greifswald, 17487 Greifswald, Germany}
\affiliation{National Superconducting Cyclotron Laboratory, East Lansing, Michigan 48824, USA}

\author{K. Gulyuz}
\affiliation{National Superconducting Cyclotron Laboratory, East Lansing, Michigan 48824, USA}

\author{A. Hamaker}
\affiliation{Department of Physics and Astronomy, Michigan State University, East Lansing, Michigan 48824, USA}
\affiliation{National Superconducting Cyclotron Laboratory, East Lansing, Michigan 48824, USA}

\author{C. Izzo}
\affiliation{Department of Physics and Astronomy, Michigan State University, East Lansing, Michigan 48824, USA}
\affiliation{National Superconducting Cyclotron Laboratory, East Lansing, Michigan 48824, USA}

\author{W.-J. Ong}
\affiliation{Department of Physics and Astronomy, Michigan State University, East Lansing, Michigan 48824, USA}
\affiliation{National Superconducting Cyclotron Laboratory, East Lansing, Michigan 48824, USA}

\author{D. Puentes}
\affiliation{Department of Physics and Astronomy, Michigan State University, East Lansing, Michigan 48824, USA}
\affiliation{National Superconducting Cyclotron Laboratory, East Lansing, Michigan 48824, USA}

\author{M. Redshaw}
\affiliation{National Superconducting Cyclotron Laboratory, East Lansing, Michigan 48824, USA}
\affiliation{Department of Physics, Central Michigan University, Mount Pleasant, Michigan 48859, USA}
\affiliation{Science of Advanced Materials Program, Central Michigan University, Mount Pleasant, Michigan 48859, USA}

\author{R. Ringle}
\affiliation{National Superconducting Cyclotron Laboratory, East Lansing, Michigan 48824, USA}

\author{R. Sandler}
\affiliation{Department of Physics, Central Michigan University, Mount Pleasant, Michigan 48859, USA}
\affiliation{Science of Advanced Materials Program, Central Michigan University, Mount Pleasant, Michigan 48859, USA}

\author{S.~Schwarz}
\affiliation{National Superconducting Cyclotron Laboratory, East Lansing, Michigan 48824, USA}

\author{C.S. Sumithrarachchi}
\affiliation{National Superconducting Cyclotron Laboratory, East Lansing, Michigan 48824, USA}

\author{J. Surbrook}
\affiliation{Department of Physics and Astronomy, Michigan State University, East Lansing, Michigan 48824, USA}
\affiliation{National Superconducting Cyclotron Laboratory, East Lansing, Michigan 48824, USA}

\author{A.C.C. Villari}
\affiliation{Facility for Rare Isotope Beams, East Lansing, Michigan 48824, USA}

\author{I.T. Yandow}
\affiliation{Department of Physics and Astronomy, Michigan State University, East Lansing, Michigan 48824, USA}
\affiliation{National Superconducting Cyclotron Laboratory, East Lansing, Michigan 48824, USA}

\date{\today}

\begin{abstract}
We report the mass measurement of $^{56}$Cu, using the LEBIT 9.4T Penning trap mass spectrometer at the National Superconducting Cyclotron Laboratory at Michigan State University. The mass of $^{56}$Cu is critical for constraining the reaction rates of the $^{55}$Ni(p,$\gamma$)$^{56}$Cu(p,$\gamma$)$^{57}$Zn($\beta^+$)$^{57}$Cu bypass around the $^{56}$Ni waiting point. Previous recommended mass excess values have disagreed by several hundred keV. Our new value, ME=$-38 626.7(6.4)$ keV, is a factor of 30 more precise than the suggested value from the 2012 atomic mass evaluation [Chin. Phys. C {\bf{36}}, 1603 (2012)], and more than a factor of 12 more precise than values calculated using local mass extrapolations, while agreeing with the newest 2016 atomic mass evaluation value  [Chin. Phys. C {\bf{41}}, 030003 (2017)]. The new experimental average was used to calculate the astrophysical $^{55}$Ni(p,$\gamma$) and $^{57}$Zn($\gamma$,p) reaction rates and perform reaction network calculations of the rp-process. These show that the rp-process flow redirects around the $^{56}$Ni waiting point through the $^{55}$Ni(p,$\gamma$) route, allowing it to proceed to higher masses more quickly and resulting in a reduction in ashes around this waiting point and an enhancement to higher-mass ashes.
\end{abstract}

\maketitle
%Introduction
Type I X-ray bursts are astronomical events that occur in binary systems where a neutron star accretes hydrogen and helium-rich material from its companion star; the accretion of more matter on the surface of the neutron star results in increasing densities and temperatures until the accreted material undergoes a thermonuclear runaway \cite{Woosley76}. The energy generated during this thermonuclear runaway gives rise to an increase in temperature and sharp increase of X-ray luminosity followed by a slower decay as the atmosphere cools. 

The high temperatures and densities achieved during this event provide the conditions necessary to trigger the rapid proton capture (rp) process, a production mechanism for proton-rich nuclei beyond the iron peak and lighter than $A\sim106$ \cite{Wallace81,Schatz01}. The rp-process flows through a series of proton capture (p,$\gamma$), photodisintegration ($\gamma$,p), $\alpha$ capture ($\alpha$,p)  and $\beta^+$-decay reactions, with relative rates of reactions determining the pathway. Type I X-ray bursts generally have rise times of $\sim1$-10 s, and decay times ranging from 10 s to several minutes, though much longer-lived superbursts, with hour-long decay times, also exist \cite{Parikh13}.Bottlenecks in the rp-process reaction pathway are created where the low proton-capture $Q$ values make photodisintegration competitive with proton-capture and $\beta^+$ decays become the dominant route. Where the $\beta^+$ decay half-life is long, relative to the timescale of the X-ray burst, a waiting point occurs. The interplay of various other factors also affects the reaction flow and thus the significance of these waiting points. Of particular importance is the relative intensity of the (p,$\gamma$) and ($\gamma$,p) reaction rates, which is highly sensitive to the $Q$ values of the reactions \cite{Parikh09}.

With a small $Q$ value for the $^{56}\text{Ni}(\text{p},\gamma)$ reaction of $Q_{\text{p},\gamma}=690.3(4)$ keV \cite{AME16} and an hours-long stellar half-life  \cite{Fuller82}, the doubly-magic nucleus $^{56}$Ni is one of the most important rp-process waiting points \cite{Kankainen10}. Indeed, it was historically thought to be the endpoint of the rp-process \cite{Wallace81}, though we now know it proceeds to higher masses \cite{Schatz01,Elomaa09}. The  flow through $^{56}$Ni is well-characterized, based on $S_p$ values \cite{Kankainen10,AME16}, as well as $^{56}$Ni(p,$\gamma$) \cite{Rehm98} and $^{57}$Cu(p,$\gamma$) \cite{Langer14} reaction rates. A route starting at $^{55}$Ni could allow rp-process flow to bypass the $^{56}$Ni waiting point through $^{55}$Ni(p,$\gamma$)$^{56}$Cu(p,$\gamma$)$^{57}$Zn($\beta^+$)$^{57}$Cu but it is not as well characterized; the branching of the flow at $^{55}$Ni between the two routes is determined by the $\beta^+$ decay rate and the $^{55}$Ni(p,$\gamma$) and $^{56}$Cu($\gamma$,p) reaction rates.

The astrophysical rates of these (p,$\gamma$) reactions can be approximated by \cite{Iliadis07}:
\begin{equation}
N_A\langle\sigma\nu\rangle \propto \sum_i(\omega\gamma)_i\exp\left(-E_i/kT\right)\label{eq:RateEq} 
\end{equation}
where $E_i=E_i^x-Q$ is the $i$th resonance for excitation energy $E^x_i$, $Q$ is the $Q$ value of the reaction, the difference in mass between the initial and final states, and $(\omega\gamma)_i$ is the $i$th resonance strength, determined by:
\begin{equation}
(\omega\gamma)_i = \frac{2J_i+1}{(2J_p+1)(2J_T+1)}\frac{\Gamma_p\Gamma_\gamma}{\Gamma_p+\Gamma_\gamma}\label{eq:resstre}
\end{equation}
where $J_i$ ,$J_p$ and $J_T$ are the spins of the resonance, proton, and ground-state proton-capturing nucleus, respectively, and $\Gamma_\gamma$ and $\Gamma_p$ are the $\gamma$ and proton partial widths. Recently, the low-lying level scheme of $^{56}$Cu was experimentally determined for the first time \cite{Ong17}, leaving the largest source of uncertainty in the critical $^{55}$Ni(p,$\gamma$) rate to be the proton separation energy of $^{56}$Cu.

Because of its high astrophysical importance, several predictions of the $^{56}$Cu atomic mass have been made recently using the Coulomb Displacement Energy (CDE) mass relation \cite{Tu16}, and the Isobaric Mass Multiplet Equation (IMME) \cite{Ong17}. Furthermore, the Atomic Mass Evaluation (\textsc{Ame}) predictions varied by several hundreds of keV from \textsc{Ame2003} \cite{AME03} to \textsc{Ame2012} \cite{AME12}. Moreover, for reaction network calculations, the masses of rp-process nuclei must be measured accurately to within 10 keV\cite{Schatz06}, a precision which is not achieved by any of the current predictions. The recently released \textsc{Ame2016} includes an unpublished atomic mass from a private communication with P. Zhang \emph{et al.}\cite{AME16} which also fails to achieve the necessary precision. Hence, we performed a high-precision mass measurement of $^{56}$Cu using Penning trap mass spectrometry, the most accurate available technique, to confirm the accuracy of that value while attaining the precision necessary for reaction network calculations to determine the flow of the rp-process around $^{56}$Ni.
%Methods and Results

In this Letter, we report the first Penning trap mass measurement of $^{56}$Cu, produced at the National Superconducting Cyclotron Laboratory (NSCL) and measured at the Low-Energy Beam and Ion Trap (LEBIT) facility \cite{Ringle13}. The LEBIT facility is unique among Penning trap mass spectrometry facilities in its ability to perform high-precision mass measurements on rare isotopes produced by projectile fragmentation. In this experiment, radioactive $^{56}$Cu was produced by impinging a 160 MeV/u primary beam of $^{58}$Ni on a 752 mg/cm$^2$ beryllium target at the Coupled Cyclotron Facility at the NSCL. The resulting beam passed through the A1900 fragment separator with a 294 mg/cm$^2$ aluminum wedge \cite{Morrissey03} to separate the secondary beam. This beam consisted of $^{56}$Cu (2.6\%), with contaminants of $^{55}$Ni, $^{54}$Co, and $^{53}$Mn.

The beam then entered the beam stopping area \cite{Cooper14} through a momentum compression beamline, where it was degraded with aluminum degraders of 205 $\mu$m and 523 $\mu$m thickness before passing through a 1010 $\mu$m, 3.1 mrad aluminum wedge and entering the gas cell with an energy of less than 1 MeV/u. In the gas cell, ions were stopped through their collision with the high-purity helium gas at a pressure of about 73 mbar; during this process, the highly-charged ions recombined down to a singly charged state. These ions were transported by a combination of RF and DC fields as well as gas flow through the gas cell, and were then extracted into a radiofrequency quadrupole (RFQ) ion-guide and transported through a magnetic dipole mass separator with a resolving power greater than $500$. Transmitted activity after the mass filter was measured using an insertable Si detector. The most activity was found with $A/Q=92$, corresponding to the extraction of $^{56}\text{Cu}$ as an adduct with two waters, $\left[^{56}\text{Cu}(\text{H}_2\text{O})_2\right]^+$. Following the mass separator, the ions then entered the LEBIT facility.

In the LEBIT facility, the $\left[^{56}\text{Cu}(\text{H}_2\text{O})_2\right]^+$ ions were first injected into the cooler-buncher, a two-staged helium-gas-filled RFQ ion trap \cite{Schwarz16}. In the first stage, moderate pressure helium gas was used to cool the ions in a large diameter RFQ ion guide. The potential difference of 55 V from the gas cell accelerated the ions into the helium gas to strip the water ligands, following the molecular-breaking technique previously used at LEBIT \cite{Schury06}. The ions were accumulated, cooled, and released to the LEBIT Penning trap in pulses of approximately 100 ns \cite{Ringle09}. To further purify the beam, a fast kicker in the beam line between the cooler-buncher and the Penning trap was used as a time-of-flight mass separator to select ions of $A/Q=56$, corresponding to $^{56}\text{Cu}^+$ and unwanted molecular contaminants of the same $A/Q$.

The 9.4T Penning trap at the LEBIT facility consists of a high-precision hyperbolic electrode system contained in an actively-shielded magnet system \cite{Ringle13}. Electrodes in front of the Penning trap are used to decelerate the ion pulses to low energy before entering the trap. The final section of these electrodes are quadrisected radially to form a ``Lorentz steerer'' \cite{Ringle07b} that forces the ion to enter the trap off-axis and perform a magnetron motion of frequency $\nu_-$ once the trapping potential is on.
\begin{figure}[t!]
\includegraphics[width=\columnwidth]{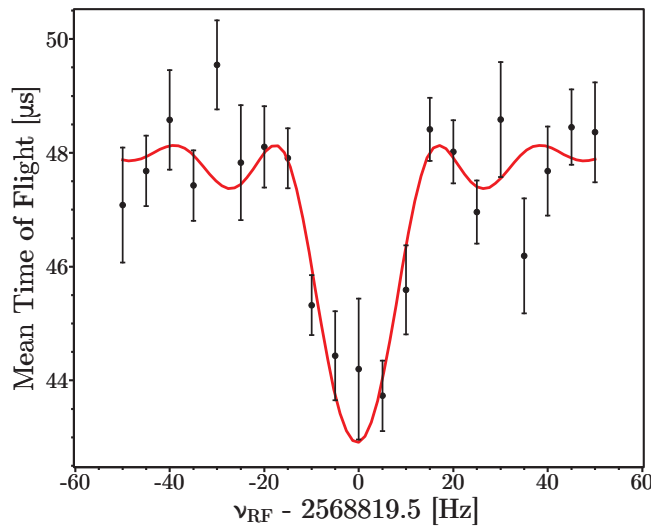}
 \caption{(color online). A sample 50-ms $^{56}\text{Cu}^+$ time-of-flight ion cyclotron resonance used for the determination of the frequency ratio of $\nu_{\text{ref}}^{\text{int}}(\text{C}_4\text{H}_7^+)/\nu_c(^{56}\text{Cu}^+)$. The solid red curve represents a fit of the theoretical profile \cite{Konig95}.\label{fig:Cu56}}
\end{figure}

After their capture, the ions were purified, using both dipole cleaning \cite{Blaum04} and the stored waveform inverse Fourier transform (SWIFT) technique \cite{Kwiatkowski15}. Both techniques excite contaminant ions using azimuthal RF dipole fields at their reduced cyclotron frequency, $\nu_+$, driving them to a large enough radius such that they do not interfere with the measurement. In the dipole technique, specific contaminants are identified for cleaning \cite{Blaum04}. In the SWIFT technique, an RF dipole drive is applied to a range of frequencies surrounding but excluding the reduced cyclotron frequency of the ion of interest, cleaning nearby contaminants without the need to specifically identify them \cite{Kwiatkowski15,Guan96}. Then, the time-of-flight ion cyclotron resonance technique (TOF-ICR) \cite{Bollen90,Konig95} was used to determine the ions' cyclotron frequency. 

In these measurements, either a 50-ms, 75-ms, or 100-ms quadrupole excitation was used. These resonances were then fitted to the theoretical line shape \cite{Konig95}, and the cyclotron frequency was thus determined; a sample 50-ms resonance of $^{56}\text{Cu}^+$ can be seen in Fig. \ref{fig:Cu56}. Between measurements of the $^{56}\text{Cu}^+$ cyclotron frequency, measurements of the reference molecular ion $\text{C}_4 \text{H}_7^+$ cyclotron frequency were conducted. The $\text{C}_4\text{H}_7$ molecule is possibly the result of an $A$ = 92 hydrocarbon molecule extracted from the gas cell and coming with the $\left[^{56}\text{Cu}(\text{H}_2\text{O})_2\right]^+$  molecule broken by collision-induced dissociation \cite{Schury06}. 

In Penning trap mass spectrometry, the experimental result is the frequency ratio \(R=\nu_{\text{ref}}^{\text{int}}/\nu_c\), where $\nu_{\text{ref}}^{\text{int}}$ is the interpolated cyclotron frequency from the $\text{C}_4 \text{H}_7^+$ measurements bracketing the $^{56}\text{Cu}^+$ measurements. Then, using the average of multiple frequency ratios $\overline{R}$ the atomic mass $M$ is given by:
\begin{equation}
M =\overline{R}\left[M_{\text{ref}}-m_e\right]+m_e,
\end{equation}\label{eq:mass}
where $M_{\text{ref}}$ is the atomic mass of the neutral reference atom or molecule, and $m_e$ the electron mass. The electron ionization energies and the molecular binding energy of $\text{C}_4\text{H}_7$, both on the order of eVs, were not included as they are several orders of magnitude smaller than the statistical uncertainty of the measurement.

\begin{figure}[b!]
\includegraphics[width=\columnwidth]{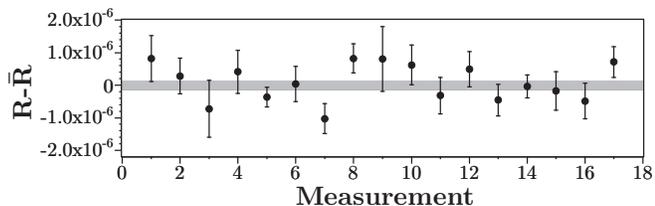}
 \caption{Measured cyclotron frequency ratios $R=\nu_{\text{ref}}^{\text{int}}/\nu_c(^{56}\text{Cu}^+)$ relative to the average value $\overline{R}$; the grey bar represents the $1\sigma$ uncertainty in $\overline{R}$.\label{fig:Cu56s}}
\end{figure}

A series of 17 measurements of the $^{56}\text{Cu}^+$ cyclotron frequency were taken over a 40-hour period and the weighted average of these measurements is $\overline{R}=$1.01641577(12). As seen in Fig. \ref{fig:Cu56s}, the individual values of $R$ scatter statistically about the average $\overline{R}$.

Most systematic uncertainties in the measured frequency ratios scale linearly with the mass difference between the ion of interest and the calibrant ion. These systematic effects include: magnetic field inhomogeneities, trap misalignment with the magnetic field, harmonic distortion of the electric potential and non-harmonic imperfections in the trapping potential \cite{Bollen90}. These mass dependent shifts to $\overline{R}$, have been studied at LEBIT and found to be at the level of $\Delta R = 2 \times 10^{-10}$/u \cite{Gulyuz15}, negligible compared to the statistical uncertainty on $\overline{R}$.

Remaining systematic effects include non-linear time-dependent changes in the magnetic field, relativistic effects on the cyclotron frequency, and ion-ion interaction in the trap. Previous work has shown that the effect of nonlinear magnetic field fluctuations on the ratio $R$ should be less than $1\times 10^{-9}$ over an hour \cite{Ringle07}, which was our measurement time. Relativistic effects on the cyclotron frequency were found to be negligible due the large mass of the ions involved. Finally, isobaric contaminants present in the trap during a measurement could lead to a systematic frequency shift \cite{Bollen92}; this effect was minimized by removing most of the contamination using the SWIFT and dipole excitations and by limiting the total number of ions in the trap. For $^{56}$Cu, the incident rate limited it to two or fewer detected ions in the trap. The $\text{C}_4 \text{H}_7$ was limited to five or fewer detected ions in the trap; a z-class analysis was performed, and any count-dependent shifts to $R$ were found to be more than an order of magnitude smaller than the statistical uncertainty.

\begin{figure}[t!]
\includegraphics[width=\columnwidth]{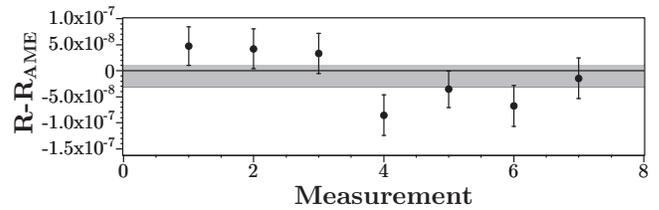}
 \caption{Difference of measured $R$ values of $^{41}\text{K}$ relative to the value calculated from \textsc{Ame2016} \cite{AME16}. The grey bar represents the average $R$ value and its $1\sigma$ uncertainty; the uncertainty of the \textsc{Ame2016} value, $1.5\times10^{-10}$, is not visible on this graph.\label{fig:Ks}}
\end{figure}

Other possible systematics unaccounted for were probed through a measurement of the ratio $R$ of stable potassium isotopes; $R=\nu_{\text{ref}}^{\text{int}}(^{39}\text{K}^+)/\nu_c(^{41}\text{K}^+)$, with SWIFT being used on the $^{41}$K measurement but not for the $^{39}$K reference, as in the experiment. Potassium was produced using the LEBIT offline thermal ion source and otherwise treated in the same way as the ions produced online. The measured $\overline{R}$ value agrees with the accepted ratio to within a Birge ratio \cite{Birge32} scaled uncertainty smaller than $2\times10^{-8}$; individual $R$ values can be seen in Fig. \ref{fig:Ks}. Thus, any mass dependent shifts either from the usage of SWIFT or the difference in mass are negligible compared to the statistical uncertainty on the $^{56}\text{Cu}$ measurement. Finally, the Birge ratio for the measurement was 1.11(12), which indicates that the fluctuations in Fig. \ref{fig:Cu56s} are statistical in nature.

The resulting mass excess is reported in Table \ref{tab:meas} as well as the recommended value from the two previous Atomic Mass Evaluations \cite{AME03,AME12}, Coulomb Displacement Energy  \cite{Tu16}, and the Isobaric Mass Multiplet Equation \cite{Ong17} predictions and the latest result from \textsc{Ame2016} \cite{AME16}. Our new $^{56}$Cu mass results in a proton separation energy of $S_p=579.8(6.4)$ keV, calculated from \(S_p(^{56}\text{Cu}) = -M(^{56}\text{Cu})+M(^{55}\text{Ni})+M(^{1}\text{H})\) using our new $^{56}$Cu mass and the masses of $^{55}$Ni and $^{1}$H from \textsc{Ame2016} \cite{AME16}. 
\begin{table}[t!]
 \caption{\label{tab:meas}A comparison of mass excesses and proton separation energies for $^{56}$Cu from CDE calculations \cite{Tu16}, IMME calculation \cite{Ong17}, and the recommended values from the last three atomic mass evaluations, and the weighted average of the two experimental measurements.}
 \begin{ruledtabular}
 \begin{tabular}{c c c}
Ref. & ME (keV) & $S_p$ (keV) \\
\hline
This work & -38 626.7(6.4) & 579.8(6.4) \\
\textsc{Ame2016} \cite{AME16} & -38 643(15) &  596(15)\\
Experimental Average & -38 629.2(5.9) & 582.3(5.9)\\
Ong \emph{et al.} \cite{Ong17} & -38 685(82) & 639(82) \\
Tu \emph{et al.} \cite{Tu16} & -38 697(88) & 651(88)\\
\textsc{Ame2003} \cite{AME03} & -38 600(140) &  560(140)\\
\textsc{Ame2012} \cite{AME12} & -38 240(200) & 190(200)
\end{tabular}
\end{ruledtabular}
\end{table}

\begin{figure}[t!]
\includegraphics[width=\columnwidth]{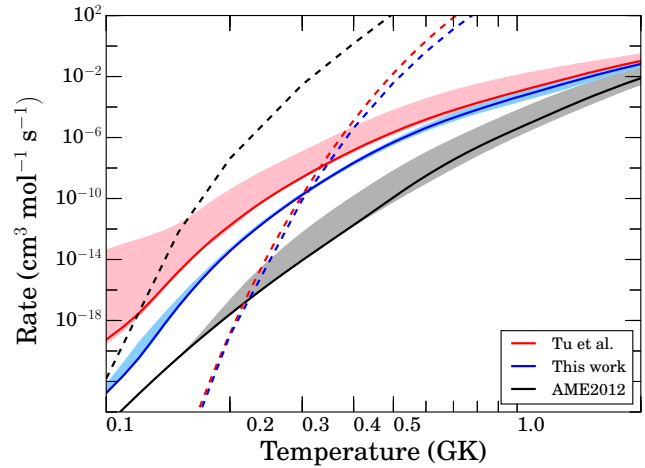}
 \caption{(color online) Rate for the $^{55}\text{Ni}(\text{p},\gamma)^{56}\text{Cu}$ reaction and 1$\sigma$ uncertainties for \textsc{Ame2012} (black band) and Tu \emph{et al.} calculated values (red band) and using our new mass measurement (blue band). The prior reverse rates (dashed lines) and new recommended reverse rate (dashed blue line) from this work are also shown. \label{fig:rateplot}}
\end{figure}

%Analysis and Conclusion
Using the weighted average of our new $^{56}$Cu mass and the \textsc{Ame16} value, also available in Table \ref{tab:meas}, and the level scheme established in \cite{Ong17}, a new astrophysical reaction rate for $^{55}\text{Ni}(\text{p},\gamma)$ was calculated. The proton and $\gamma$ widths, $\Gamma_\text{p}$ and $\Gamma_{\gamma}$, were calculated for each state using a shell model with the GXPF1A interaction \cite{Honma05}. Up to three-particle-three-hole excitations in the $pf$ shell were allowed in this calculation \cite{Ong17}, with the proton and $\gamma$ widths and resonance strengths scaled appropriately. A Monte Carlo approach, similar to that in \cite{Iliadis15,Ong17}, was used to calculate uncertainties. At a given temperature, the 50th percentile of the distribution of calculated rates gives the median rate, and the 16th and 84th percentiles the 1$\sigma$ uncertainties. The results can be seen in Fig. \ref{fig:rateplot}, compared with the results found using the extrema of the calculated $^{56}$Cu masses, \textsc{Ame2012}\cite{AME12} and Tu \emph{et al.} \cite{Tu16}; this shows that the (p,$\gamma$) reaction dominates up to $\sim 0.3$ GK, slightly lower than the Tu \emph{et al.} case, and significantly higher than the  \textsc{Ame2012} case, where the reverse rate always dominates. For the \textsc{Ame2012} mass, at low temperatures, direct capture dominates, leading to little uncertainty, but at higher temperatures, the reaction can access resonant states and the mass uncertainty dominates. The uncertainty of the rate at low temperatures for the Tu \emph{et al.} value is dominated by the potential of the 573 keV state to act as a resonance within the 1-$\sigma$ lower bound of the $Q$ value; this is also why the median rate is near the lower bound. Our mass shows a reduced uncertainty when compared to both prior masses, as the $Q$ value uncertainty is now comparable to the uncertainty in resonance energies.
\begin{figure}[b!]
\includegraphics[width=\columnwidth]{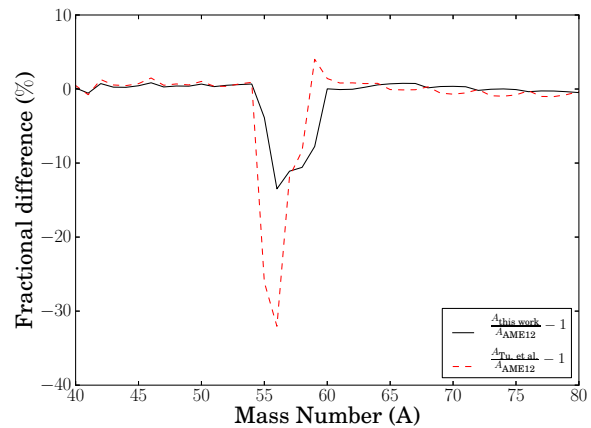}
 \caption{(color online) Fractional difference of abundance by mass number of this work (solid black) compared to that using the masses suggested, in \textsc{Ame2012}\cite{AME12} and the same fractional difference using Tu \emph{et al.}\cite{Tu16} (dashed red). \label{fig:abunplot}}
\end{figure}

A single-zone X-ray burst model was then run using the new $^{56}$Cu mass with an ignition temperature of 0.386~GK, ignition pressure of 1.73 $\times$ 10$^{22}$ erg cm$^{-3}$ and initial hydrogen and helium mass fractions of 0.51 and 0.39 respectively, demonstrated by \cite{Cyburt16} to produce light curves and ash compositions to most closely match those of multi-zone models, and with a peak temperature of 1.17 GK. As can be seen in Fig. \ref{fig:abunplot}, the final abundances produced by this calculation demonstrate the extent to which the bypass due to the change in (p,$\gamma$)-($\gamma$,p) equilibrium is active, showing a reduction in abundance in the mass range around the $^{56}$Ni waiting point in comparison to ones based on the suggested \textsc{Ame2012} value, though not as extreme as the one seen with the mass from Tu \emph{et al.}; the maximal bypass is 39\%, with a typical X-ray burst trajectory having a bypass of 15\%. The percentage increase in heavier mass ashes is not as apparent due to the higher absolute abundance of heavier ashes at around mass 60. This means the newly-calculated reaction rate allows the rp-process flow to bypass the waiting point and proceed more quickly through the region. 

In summary, the high precision measurement of the mass of $^{56}$Cu is reported, allowing the calculation of its proton separation energy to a precision of 6.5 keV, a factor of 30 improvement over the \textsc{Ame2012} suggested value and a factor of more than 12 improvement over the IMME and CDE calculated values \cite{Ong17,Tu16} while agreeing with the private communication available in \textsc{Ame2016} \cite{AME16}. New thermonuclear reaction rates were then calculated using the first experimental mass of $^{56}$Cu, the weighted average of our new value and the \textsc{Ame2016} value, and abundances for the rp-process around the $^{56}$Ni waiting point were determined. These abundances show that the new reaction rate allows the rp-process to redirect around this waiting point and proceed to heavier masses more quickly, resulting in an enhancement in higher-mass ashes. The dominant sources of uncertainty are now the unmeasured widths $\Gamma_{\text{p}}$ and $\Gamma_\gamma$ for the $^{55}$Ni(p,$\gamma$) reaction; the unmeasured higher-lying level scheme of $^{56}$Cu; the unmeasured $^{57}$Zn mass for the $^{56}$Cu(p,$\gamma$) and $^{57}$Zn($\gamma$,p) reactions, which hampers this flow from bypassing $^{56}$Ni at high temperatures; and the high uncertainty on the $\beta$-delayed proton branch of $^{57}$Zn (78(17)\%, \cite{Blank07}), which directs flow back to $^{56}$Ni.

The authors would like to acknowledge Hendrik Schatz for fruitful discussions related to the work presented in this Letter. This work was conducted with the support of Michigan State University, the National Science Foundation under Contract No. PHY-1102511 and No. PHY-1713857,  and the US Department of Energy, Office of Science, Office of Nuclear Physics under award number DE-SC0015927. The work leading to this publication has also been supported by a DAAD P.R.I.M.E. fellowship with funding from the German Federal Ministry of Education and Research and the People Programme (Marie Curie Actions) of the European Union’s Seventh Framework Programme FP7/2007/2013) under REA Grant Agreement No. 605728.

\end{document}